\pdfoutput=1

\documentclass[12pt,a4paper]{article}

\usepackage{ifthen} 
\newboolean{pdflatex}
\setboolean{pdflatex}{true} 

\newboolean{articletitles}
\setboolean{articletitles}{true} 

\newboolean{uprightparticles}
\setboolean{uprightparticles}{false} 

\usepackage{longtable} 

\usepackage{microtype}
\usepackage{lineno}  
\usepackage{xspace} 
\usepackage{caption} 

\usepackage{graphicx}  
\usepackage{color}
\usepackage{colortbl}
\graphicspath{{./figs/}} 

\usepackage{amsmath} 
\usepackage{amssymb}
\usepackage{amsfonts}
\usepackage{upgreek} 

\newcommand*\patchAmsMathEnvironmentForLineno[1]{%
\expandafter\let\csname old#1\expandafter\endcsname\csname #1\endcsname
\expandafter\let\csname oldend#1\expandafter\endcsname\csname
end#1\endcsname
 \renewenvironment{#1}%
   {\linenomath\csname old#1\endcsname}%
   {\csname oldend#1\endcsname\endlinenomath}%
}
\newcommand*\patchBothAmsMathEnvironmentsForLineno[1]{%
  \patchAmsMathEnvironmentForLineno{#1}%
  \patchAmsMathEnvironmentForLineno{#1*}%
}
\AtBeginDocument{%
\patchBothAmsMathEnvironmentsForLineno{equation}%
\patchBothAmsMathEnvironmentsForLineno{align}%
\patchBothAmsMathEnvironmentsForLineno{flalign}%
\patchBothAmsMathEnvironmentsForLineno{alignat}%
\patchBothAmsMathEnvironmentsForLineno{gather}%
\patchBothAmsMathEnvironmentsForLineno{multline}%
\patchBothAmsMathEnvironmentsForLineno{eqnarray}%
}

\usepackage{hyperref}    
\usepackage[all]{hypcap} 

\usepackage{xspace} 
\usepackage{upgreek}

\def\lhcb {\mbox{LHCb}\xspace}

\def\MagUp {\mbox{\em Mag\kern -0.05em Up}\xspace}

\ifthenelse{\boolean{uprightparticles}}%
{

 \def\Ppsi        {\ensuremath{\uppsi}\xspace}

 \def\PDelta      {\ensuremath{\Delta}\xspace}                 
 \def\PXi      {\ensuremath{\Xi}\xspace}                 
 \def\PLambda      {\ensuremath{\Lambda}\xspace}                 
 \def\PSigma      {\ensuremath{\Sigma}\xspace}                 
 \def\POmega      {\ensuremath{\Omega}\xspace}                 
 \def\PUpsilon      {\ensuremath{\Upsilon}\xspace}                 
 
 \def\PB      {\ensuremath{\mathrm{B}}\xspace}                 
                  
 \def\PD      {\ensuremath{\mathrm{D}}\xspace}

 \def\PJ      {\ensuremath{\mathrm{J}}\xspace}                 
 \def\PK      {\ensuremath{\mathrm{K}}\xspace}

 \def\Pb      {\ensuremath{\mathrm{b}}\xspace}

 \def\Pi      {\ensuremath{\mathrm{i}}\xspace}

}
{

 \def\Ppsi        {\ensuremath{\psi}\xspace}                 
                  
 \mathchardef\PDelta="7101
 \mathchardef\PXi="7104
 \mathchardef\PLambda="7103
 \mathchardef\PSigma="7106
 \mathchardef\POmega="710A
 \mathchardef\PUpsilon="7107
                  
 \def\PB      {\ensuremath{B}\xspace}                 
                  
 \def\PD      {\ensuremath{D}\xspace}

 \def\PJ      {\ensuremath{J}\xspace}                 
 \def\PK      {\ensuremath{K}\xspace}

 \def\Pb      {\ensuremath{b}\xspace}

 \def\Pi      {\ensuremath{i}\xspace}

}

\makeatletter
  \newcommand{\miniscule}{\@setfontsize\miniscule{5}{6}}
\makeatother

\DeclareRobustCommand{\optbar}[1]{\shortstack{{\miniscule (\rule[.5ex]{1.25em}{.18mm})}
  \\ [-.7ex] $#1$}}

\def\H      {{\ensuremath{\PH^0}}\xspace}

\def\bquark    {{\ensuremath{\Pb}}\xspace}

\def\kaon    {{\ensuremath{\PK}}\xspace}
  \def\Kbar    {{\kern 0.2em\overline{\kern -0.2em \PK}{}}\xspace}

\def\KorKbar    {\kern 0.18em\optbar{\kern -0.18em K}{}\xspace}

\def\Km      {{\ensuremath{\kaon^-}}\xspace}

  \def\Dbar    {{\kern 0.2em\overline{\kern -0.2em \PD}{}}\xspace}

\def\DorDbar    {\kern 0.18em\optbar{\kern -0.18em D}{}\xspace}

\def\Bbar    {{\ensuremath{\kern 0.18em\overline{\kern -0.18em \PB}{}}}\xspace}

\def\BorBbar    {\kern 0.18em\optbar{\kern -0.18em B}{}\xspace}

\def\jpsi     {{\ensuremath{{\PJ\mskip -3mu/\mskip -2mu\Ppsi\mskip 2mu}}}\xspace}

  \def\Y#1S{\ensuremath{\PUpsilon{(#1S)}}\xspace}

\def\Lz          {{\ensuremath{\PLambda}}\xspace}
\def\Lbar        {{\ensuremath{\kern 0.1em\overline{\kern -0.1em\PLambda}}}\xspace}
\def\LorLbar    {\kern 0.18em\optbar{\kern -0.18em \PLambda}{}\xspace}

\def\Sigmares    {{\ensuremath{\PSigma}}\xspace}

\def\Lb      {{\ensuremath{\Lz^0_\bquark}}\xspace}

\def\to                 {\ensuremath{\rightarrow}\xspace}

\def\AT#1     {\ensuremath{A_{\mathrm{T}}^{#1}}\xspace}           

\def\C#1      {\ensuremath{\mathcal{C}_{#1}}\xspace}                       
\def\Cp#1     {\ensuremath{\mathcal{C}_{#1}^{'}}\xspace}                    
\def\Ceff#1   {\ensuremath{\mathcal{C}_{#1}^{\mathrm{(eff)}}}\xspace}        
\def\Cpeff#1  {\ensuremath{\mathcal{C}_{#1}^{'\mathrm{(eff)}}}\xspace}       
\def\Ope#1    {\ensuremath{\mathcal{O}_{#1}}\xspace}                       
\def\Opep#1   {\ensuremath{\mathcal{O}_{#1}^{'}}\xspace}                    

\newcommand{\tev}{\ifthenelse{\boolean{inbibliography}}{\ensuremath{~T\kern -0.05em eV}\xspace}{\ensuremath{\mathrm{\,Te\kern -0.1em V}}}\xspace}
\newcommand{\gev}{\ensuremath{\mathrm{\,Ge\kern -0.1em V}}\xspace}
\newcommand{\mev}{\ensuremath{\mathrm{\,Me\kern -0.1em V}}\xspace}
\newcommand{\kev}{\ensuremath{\mathrm{\,ke\kern -0.1em V}}\xspace}
\newcommand{\ev}{\ensuremath{\mathrm{\,e\kern -0.1em V}}\xspace}
\newcommand{\gevc}{\ensuremath{{\mathrm{\,Ge\kern -0.1em V\!/}c}}\xspace}
\newcommand{\mevc}{\ensuremath{{\mathrm{\,Me\kern -0.1em V\!/}c}}\xspace}
\newcommand{\gevcc}{\ensuremath{{\mathrm{\,Ge\kern -0.1em V\!/}c^2}}\xspace}
\newcommand{\gevgevcccc}{\ensuremath{{\mathrm{\,Ge\kern -0.1em V^2\!/}c^4}}\xspace}
\newcommand{\mevcc}{\ensuremath{{\mathrm{\,Me\kern -0.1em V\!/}c^2}}\xspace}

\def\invfb   {\ensuremath{\mbox{\,fb}^{-1}}\xspace}

\def\gsim{{~\raise.15em\hbox{$>$}\kern-.85em
          \lower.35em\hbox{$\sim$}~}\xspace}
\def\lsim{{~\raise.15em\hbox{$<$}\kern-.85em
          \lower.35em\hbox{$\sim$}~}\xspace}

\def\PDF {PDF\xspace}

\def\sPlot{\mbox{\em sPlot}\xspace}

\def\tell1  {TELL1\xspace}
\def\ukl1   {UKL1\xspace}

\usepackage{cite} 
\usepackage{mciteplus}

\begin{document}

\renewcommand{\thefootnote}{\fnsymbol{footnote}}
\setcounter{footnote}{1}


\begin{titlepage}
\pagenumbering{roman}

\vspace*{-1.5cm}
\centerline{\large EUROPEAN ORGANIZATION FOR NUCLEAR RESEARCH (CERN)}
\vspace*{0.5cm}
\hspace*{-0.5cm}
\begin{tabular*}{\linewidth}{lc@{\extracolsep{\fill}}r}
\ifthenelse{\boolean{pdflatex}}
{\vspace*{-3.1cm}\mbox{\!\!\!\includegraphics[width=.14\textwidth]{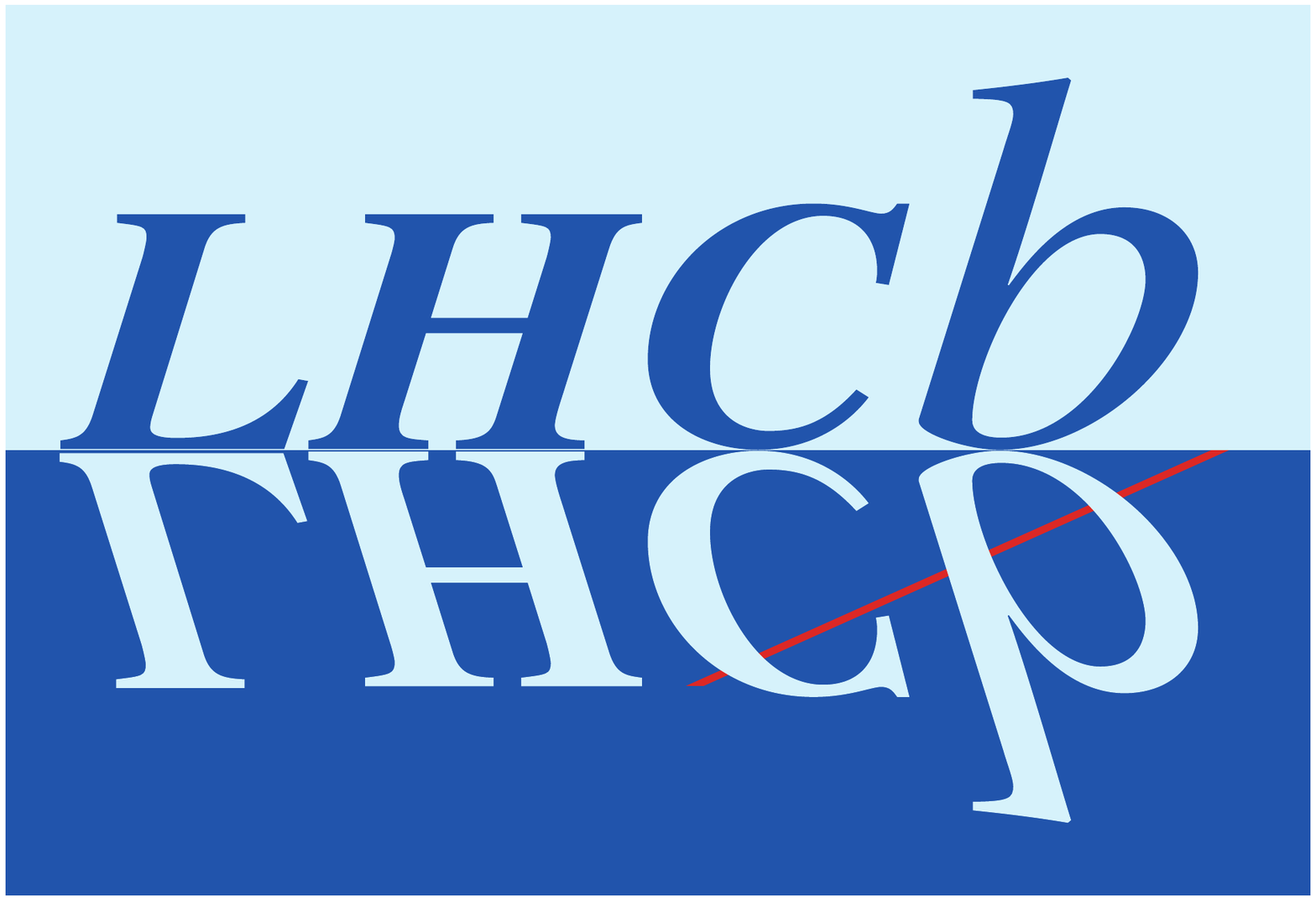}} & &}%
{\vspace*{-1.2cm}\mbox{\!\!\!\includegraphics[width=.12\textwidth]{lhcb-logo.eps}} & &}%
\\
 & & CERN-EP-2016-086 \\  
 & & LHCb-PAPER-2016-009 \\  
 & & April 19, 2016 \\ 
 & & \\
\end{tabular*}

\vspace*{4.0cm}

{\bf\boldmath\huge
\begin{center}
Model-independent evidence for $J/\psi p$ contributions to $\Lb\to J/\psi p K^-$ decays  
\end{center}
}

\vspace*{1.0cm}

\begin{center}
The LHCb collaboration\footnote{Authors are listed at the end of this paper.}
\end{center}

\vspace{\fill}

\begin{abstract}
  \noindent
The data sample of $\Lb\to\jpsi p K^-$ decays acquired with the LHCb detector 
from 7 and 8~TeV $pp$ collisions, corresponding to an integrated luminosity of 3~\invfb,
is inspected for the presence of $\jpsi p$ or $\jpsi K^-$ contributions  
with minimal assumptions about $K^- p$ contributions.
It is demonstrated at more than 9 standard deviations 
that $\Lb\to\jpsi p K^-$ decays cannot be described with $K^- p$ contributions
alone, and that $\jpsi p$ contributions play a dominant role in this incompatibility.
These model-independent results support the previously obtained model-dependent 
evidence for $P_c^+\to\jpsi p$ charmonium-pentaquark states in the same data sample.

\end{abstract}

\vspace*{1.0cm}

\begin{center}
  Submitted to Physical Review Letters
\end{center}

\vspace{\fill}

{\footnotesize 
\centerline{\copyright~CERN on behalf of the \lhcb collaboration, license \href{http://creativecommons.org/licenses/by/4.0/
}{CC-BY-4.0}.}}
\vspace*{2mm}

\end{titlepage}


\newpage
\setcounter{page}{2}
\mbox{~}
\newpage


\cleardoublepage


\renewcommand{\thefootnote}{\arabic{footnote}}
\setcounter{footnote}{0}



\pagestyle{plain} 
\setcounter{page}{1}
\pagenumbering{arabic}


%

\newboolean{prl}
\setboolean{prl}{false} 

\newboolean{supp}
\setboolean{supp}{false} 

\newlength{\figsize}
\setlength{\figsize}{0.9\hsize}
\def\PDF{{\cal F}}
\def\ZP{P_c}
\def\LambdaStar{{\Lz^*}}
\def\LambdaStarn{{\Lz^*_{\!n}}}
\def\H{{\cal H}}
\def\F#1{\{#1\}}
\def\BA#1#2#3{{#1}_{{#2}}^{\,\,\F{\!#3\!}}}
\def\lmax{l_{\rm max}}
\def\llarge{l_{\rm large}}
\def\pschi{\chi}
\def\ndf{{\rm ndf}}
\def\dll{{\Delta(\!-2\ln L\!)}}

\noindent
From the birth of the quark model, 
it has been anticipated that baryons could be constructed not only from three quarks, 
but also from four quarks and an antiquark \cite{GellMann:1964nj,Zweig:1964}, 
hereafter referred to as pentaquarks. 
The distribution of $\jpsi p$ mass ($m_{\jpsi p}$) 
in $\Lb\to\jpsi p K^-$, $\jpsi\to\mu^+\mu^-$ 
decays observed with the LHCb detector at the LHC shows 
a narrow peak suggestive of $uudc\bar c$ pentaquark formation,
amidst the dominant formation of various excitations 
of the $\Lz$ $[uds]$ baryon ($\LambdaStar$) decaying to $K^-p$ \cite{LHCb-PAPER-2015-029}. 
(The inclusion of charge conjugate states is implied in this Letter.)
Amplitude analyses were performed on all relevant masses and decay 
angles of the six-dimensional (6D) data, using the helicity formalism and 
Breit-Wigner amplitudes to describe all resonances. In addition to the 
previously well established $\LambdaStar$ resonances, two pentaquark 
resonances $P_c(4380)^+$ ($9\,\sigma$ significance) and $P_c(4450)^+$ ($12\,\sigma$) 
were required in the model for a good description of the data.
The mass, width and fit fractions were determined 
to be $4380\pm8\pm29$\mev, $205\pm18\pm86$\mev, $(8.4\pm0.7\pm4.3)\%$,  
and $4450\pm2\pm3$\mev, $39\pm5\pm19$\mev, $(4.1\pm0.5\pm1.1)\%$, 
respectively. 

The addition of further $\LambdaStar$ states beyond the well-established ones, and 
of nonresonant contributions, did not remove the need for two pentaquark states
in the model to describe the data.
Yet $\LambdaStar$ spectroscopy is a complex problem, as pointed out in a
 recent reanalysis of $\overline{K}N$ scattering data \cite{SPECRDMMS},   
in which the well-established $\Lz(1800)$ state was not seen, 
and evidence for a few previously unidentified states was obtained. 
Theoretical models of $\LambdaStar$ baryons \cite{SPECFG, SPECCI, SPECLMP, SPECMPS, SPECSF, SPECELMS} 
predict a much larger number of higher mass excitations than is established experimentally \cite{PDG2014}. 
The high density of predicted states, presumably with large widths, 
would make it difficult to identify them experimentally. 
Nonresonant contributions with non-trivial $K^-p$ mass-dependence may also be present. 
Therefore, it is worth inspecting the $\Lb\to\jpsi p K^-$ data with 
an approach that is model-independent with respect to $K^-p$ contributions. 
Such a method was introduced by the BaBar collaboration~\cite{Aubert:2008aa} and later improved upon 
by the LHCb collaboration~\cite{LHCb-PAPER-2015-038}.
There it was used to examine 
$B^0\to\psi(2S) \pi^+K^-$ decays, 
which are dominated by kaon excitations decaying to $K^-\pi^+$, 
in order to understand whether the data require the presence of the tetraquark 
candidate decay, $Z(4430)^+\to\psi(2S)\pi^+$.  
In this Letter, this method is applied to the same $\Lb\to\jpsi p K^-$ sample previously 
analyzed in the amplitude analysis \cite{LHCb-PAPER-2015-029}.  
The sensitivity of the model-independent approach 
to exotic resonances is investigated with simulation studies.  

The \lhcb detector is a single-arm forward spectrometer covering the pseudorapidity range \mbox{$2<\eta<5$}, 
described in detail in Ref.~\cite{Alves:2008zz}.  
The data selection is described in Ref.~\cite{LHCb-PAPER-2015-029}.  
A mass window of $\pm2\,\sigma$ ($\sigma=7.5$~MeV) 
around the $\Lb$ mass peak is selected, leaving $n_{\rm cand}^{\rm sig}=27\,469$ \Lb candidates for further analysis,
with background fraction ($\beta$) equal to 5.4\%. 
The background is subtracted using  
$n_{\rm cand}^{\rm side}=10\,259$ candidates from the $\Lb$ sidebands,
which extend from $\pm38$ to $\pm140\mev$ from the peak (see the supplemental material). 

The aim of this analysis is to assess the level of consistency of the data with the hypothesis 
that all $\Lb\to\jpsi p K^-$ decays proceed via $\Lb\to\jpsi\LambdaStar$, $\LambdaStar\to p K^-$, 
with minimal assumptions about the spin and lineshape of possible $\LambdaStar$ contributions. 
This will be referred to as the null-hypothesis $H_0$. 
Here, $\LambdaStar$ denotes not only excitations of the $\Lz$ baryon, 
but also nonresonant $K^-p$ contributions or excitations of the $\Sigmares$ baryon. 
The latter contributions are expected to be small\cite{Donoghue:1979mu}.
The analysis method is two-dimensional and uses the information contained in the Dalitz variables, 
$(m_{Kp}^2,m_{\jpsi p}^2)$, or equivalently in $(m_{Kp},\cos\theta_{\LambdaStar})$,
where $\theta_{\LambdaStar}$ is the helicity angle of the $K^-p$ system, 
defined as the angle between the $\vec{p}_K$ and 
$-\vec{p}_{\Lb}$ (or $-\vec{p}_{\jpsi}$) directions in the $K^-p$ rest frame. 

\begin{figure}[tbp]
   \vskip-0.5cm
  \begin{center}
  \includegraphics*[width=\figsize]{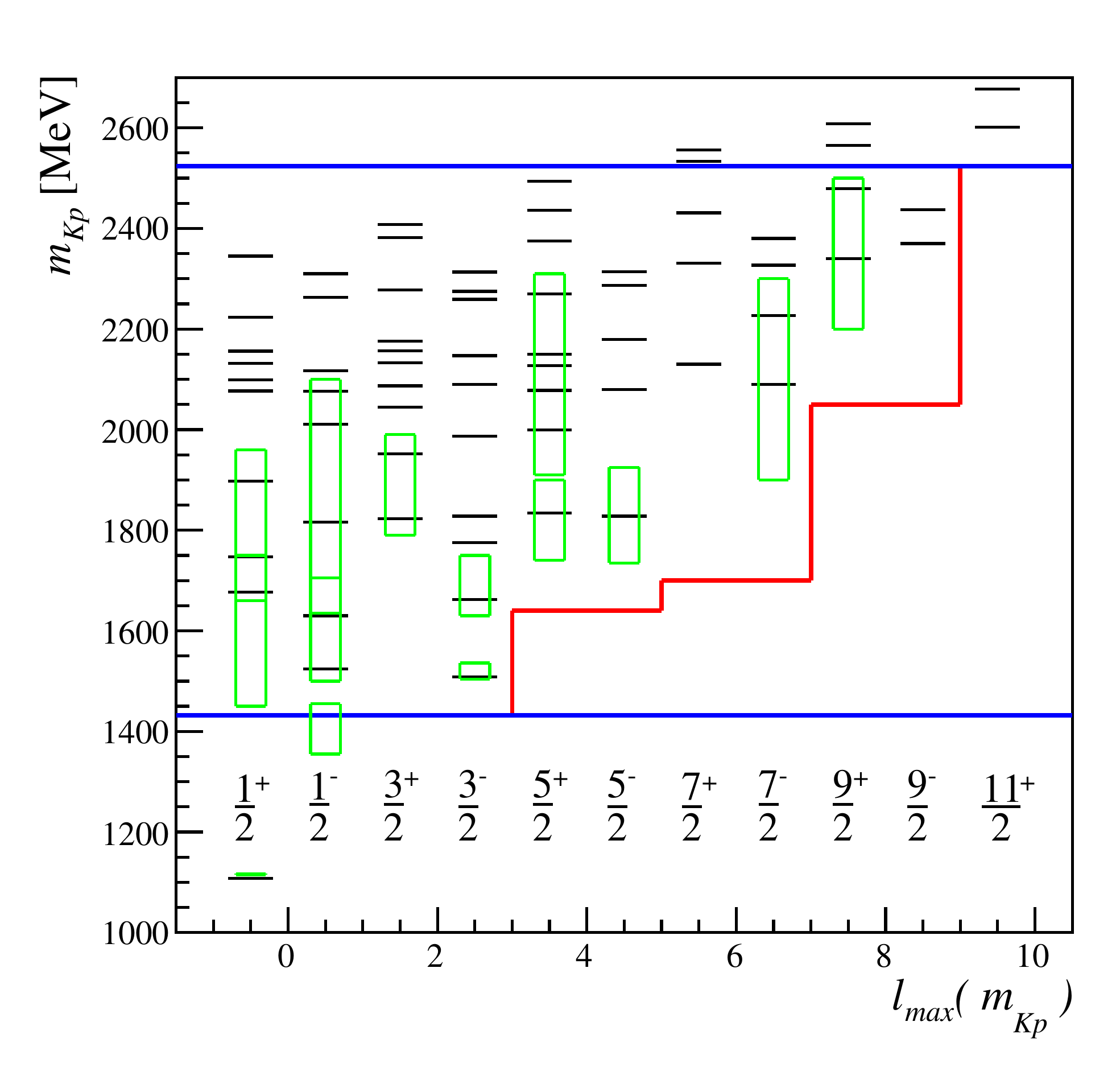}
  \end{center}
  \vskip-1.1cm\caption{\small 
   Excitations of the $\Lz$ baryon. States predicted 
   in Ref.~\cite{SPECLMP} are shown as short 
   horizontal bars (black)
   and experimentally well-established $\LambdaStar$ states 
   are shown as green
   boxes covering the mass ranges from 
   $M_0-\Gamma_0$ to $M_0+\Gamma_0$.
   The $m_{Kp}$ mass range probed in $\Lb\to\jpsi p \Km$ decays
   is shown by long horizontal lines (blue).
   The $\lmax(m_{Kp})$ filter is shown as a stepped line (red).
   All contributions from $\LambdaStar$ states with 
   $J^P$ values to the left of the red line 
   are accepted by the filter. 
   The filter works well also for the excitations of the $\Sigmares$ baryon
   \cite{PDG2014,SPECLMP} (not shown). 
   \label{fig:lmax}
  }
   \vskip-0.5cm
\end{figure}

The $(m_{Kp},\cos\theta_{\LambdaStar})$ plane
is particularly suited for implementing constraints stemming 
from the $H_0$ hypothesis by expanding the $\cos\theta_{\LambdaStar}$ angular distribution 
in Legendre polynomials $P_l$: 
\begin{equation} 
dN/d\cos\theta_{\LambdaStar}=\sum_{l=0}^{\lmax} \langle P_{l}^U \rangle P_l(\cos\theta_{\LambdaStar}),
\notag
\end{equation}
where $N$ is the efficiency-corrected and background-subtracted signal yield, and $\langle P_{l}^U \rangle$ 
is an unnormalized Legendre moment of rank $l$, 
\begin{equation}
\langle P_{l}^U \rangle= \int_{-1}^{+1} d\cos\theta_{\LambdaStar}\, P_l(\cos\theta_{\LambdaStar}) \,
dN/d\cos\theta_{\LambdaStar}.
\notag
\end{equation}
Under the $H_0$ hypothesis, $K^-p$ components cannot contribute to moments
of rank higher than $2\,J_{\rm max}$,  
where $J_{\rm max}$ is the highest spin of any $K^-p$ contribution at the given $m_{Kp}$ value.
This requirement sets the appropriate $\lmax$ value,
which can be deduced 
from the lightest experimentally known $\LambdaStar$ resonances for each $J$, or
from the quark model, as in Fig.~\ref{fig:lmax}.
An $\lmax(m_{Kp})$ function is formed,  
guided by the values of resonance masses ($M_0$) 
lowered by two units of their widths ($\Gamma_0$):
$\lmax=3$ for $m_{Kp}$ up to $1.64$ \gev,
5 up to $1.70$ \gev,
7 up to $2.05$ \gev and
9 for higher masses
as visualized in Fig.~\ref{fig:lmax}.

Reflections from other channels, 
$\Lb\to P_c^+K^-$, $P_c^+\to\jpsi p$ or $\Lb\to Z_c^- p$, $Z_c^-\to \jpsi K^-$, 
would introduce both low and high rank moments
(see the supplemental material for an illustration).
The narrower the resonance, the narrower the reflection and the higher the rank $l$ 
of Legendre polynomials required to describe such a structure.  

Selection criteria and backgrounds can also produce high-$l$ structures in the $\cos\theta_{\LambdaStar}$ distribution. 
Therefore, the data are efficiency-corrected and the background is subtracted. 
Even though testing the $H_0$ hypothesis 
involves only two dimensions, the selection efficiency has some 
dependence on the other phase-space dimensions, namely the $\Lb$ and $\jpsi$ helicity angles, 
as well as angles between the $\Lb$ decay plane and the $\jpsi$ and $\LambdaStar$
decay planes. 
Averaging the efficiency over these additional dimensions ($\Omega_a$) would introduce 
biases dependent on the exact dynamics of the $\LambdaStar$ decays. 
Therefore, a six-dimensional efficiency correction is used.  
The efficiency parameterization, $\epsilon(m_{Kp},\cos\theta_{\LambdaStar},\Omega_a)$,  
is the same as that used in the amplitude analysis and is described in Sec.~5 of the supplement 
of Ref.~\cite{LHCb-PAPER-2015-029}.

\begin{figure}[hbtp]
   \vskip-0.2cm
  \begin{center}
  \includegraphics*[width=1.05\figsize]{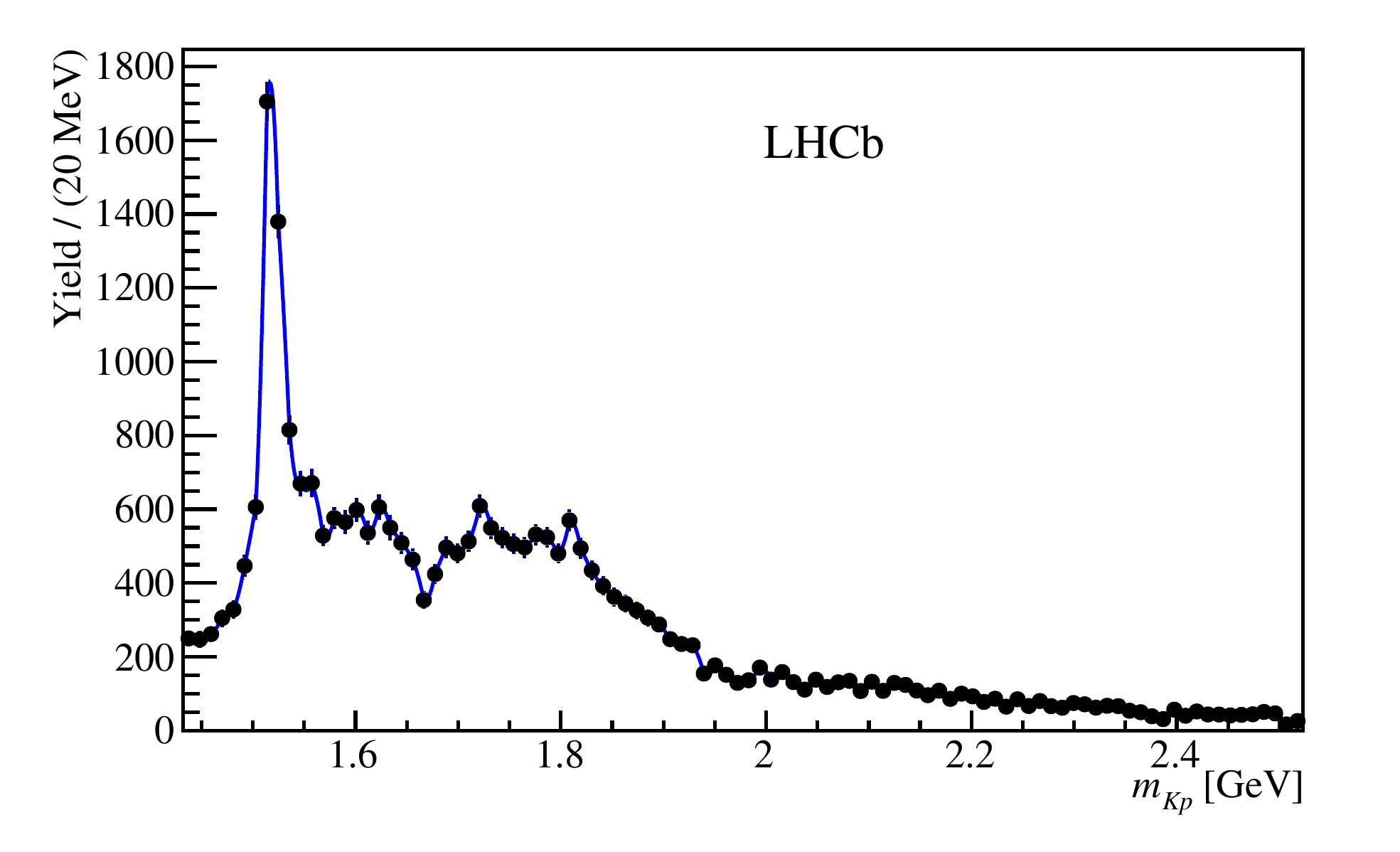}
  \end{center}
  \vskip-0.5cm\caption{\small 
    Efficiency-corrected and background-subtracted $m_{Kp}$ distribution 
    of the data (black points with error bars),
    with $\PDF(m_{Kp}|H_0)$ superimposed (solid blue line). $\PDF(m_{Kp}|H_0)$ fits 
    the data by construction.  
   \label{fig:mkp}
  }
   \vskip-0.5cm
\end{figure}

\begin{figure}[tbp]
   \vskip-0.2cm
  \begin{center}
  \includegraphics*[width=\figsize]{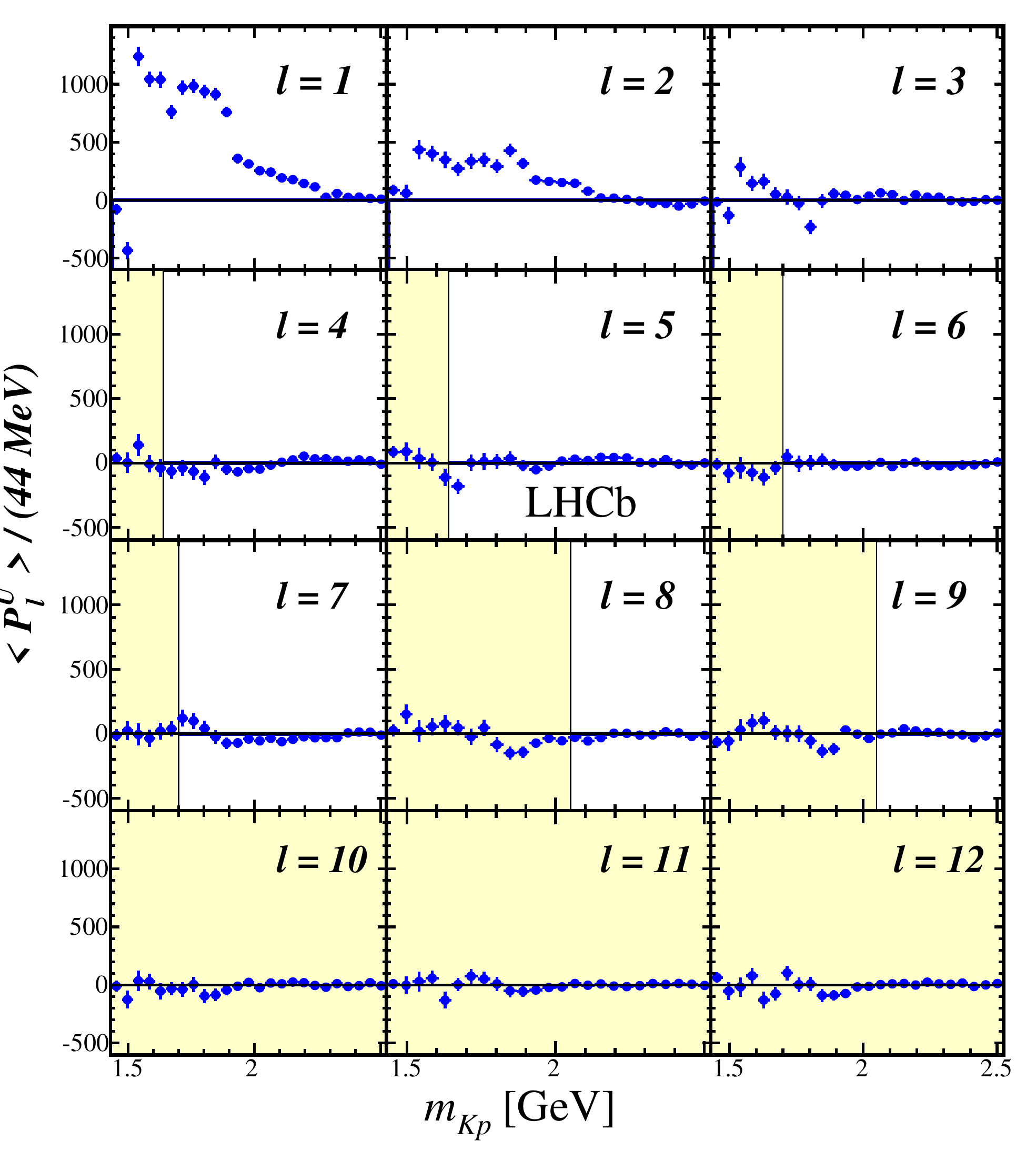}
  \end{center}
  \vskip-0.5cm\caption{\small 
           Legendre moments of $\cos\theta_{\LambdaStar}$ as a function of $m_{Kp}$ in the data.
           Regions excluded by the $l\le\lmax(m_{Kp})$ filter are shaded.
   \label{fig:moments}
  }
   \vskip-0.2cm
\end{figure}

In order to make the analysis as model-independent as possible, no interpretations
are imposed on the $m_{Kp}$ distribution. 
Instead, the observed efficiency-corrected
and background-subtracted histogram of $m_{Kp}$ is used. 
To obtain a continuous probability density function, $\PDF(m_{Kp}|H_0)$, a quadratic interpolation of 
the histogram is performed, as shown in Fig.~\ref{fig:mkp}. 
The essential part of this analysis method is to incorporate the $l\leq\lmax(m_{Kp})$ 
constraint on the $\LambdaStar$ helicity angle distribution: 
$\PDF(m_{Kp},\cos\theta_{\LambdaStar}|H_0)= \PDF(m_{Kp}|H_0)\, \PDF(\cos\theta_{\LambdaStar}|H_0, m_{Kp})$, 
where $\PDF(\cos\theta_{\LambdaStar}|H_0, m_{Kp})$ is obtained  
via linear interpolation between neighboring $m_{Kp}$ bins of  
\begin{equation}
\PDF(\cos\theta_{\LambdaStar}|H_0,{m_{Kp}}^k)=\sum_{l=0}^{\lmax({m_{Kp}}^k)} 
\langle P_{l}^N \rangle^k P_l(\cos\theta_{\LambdaStar}),
\notag
\end{equation}
where $k$ is the bin index. 
Here the Legendre moments $\langle P_{l}^N \rangle^k$
are normalized by the yield in the corresponding $m_{Kp}$ bin,
since the overall normalization of  $\PDF(\cos\theta_{\LambdaStar}|H_0, m_{Kp})$ 
to the data is already contained in the $\PDF(m_{Kp}|H_0)$ definition. 
The data are used to determine
\begin{equation}
\langle P_{l}^U \rangle^k = \sum_{i=1}^{{n_{\rm cand}}^k} (w_i/\epsilon_i)  P_l(\cos\theta_{\LambdaStar}^i). 
\notag
\end{equation}
Here the index $i$ runs over selected $\jpsi p K^-$ candidates in 
the signal and sideband regions for the $k^{th}$ bin of $m_{Kp}$  
(${n_{\rm cand}}^k$ is their total number),
$\epsilon_i=\epsilon( {m_{Kp}}^i,\cos{\theta_{\LambdaStar}}^i,{\Omega_a}^i)$ is the efficiency correction, 
and $w_i$ is the background subtraction weight, which equals 1
for events in the signal region and $- \beta\,n_{\rm cand}^{\rm sig}/n_{\rm cand}^{\rm side}$ for
events in the sideband region. 
Values of $\langle P_{l}^U \rangle^k$ are shown in Fig.~\ref{fig:moments}. 

\begin{figure}[hbtp]
  \begin{center}
  \includegraphics*[width=1.05\figsize]{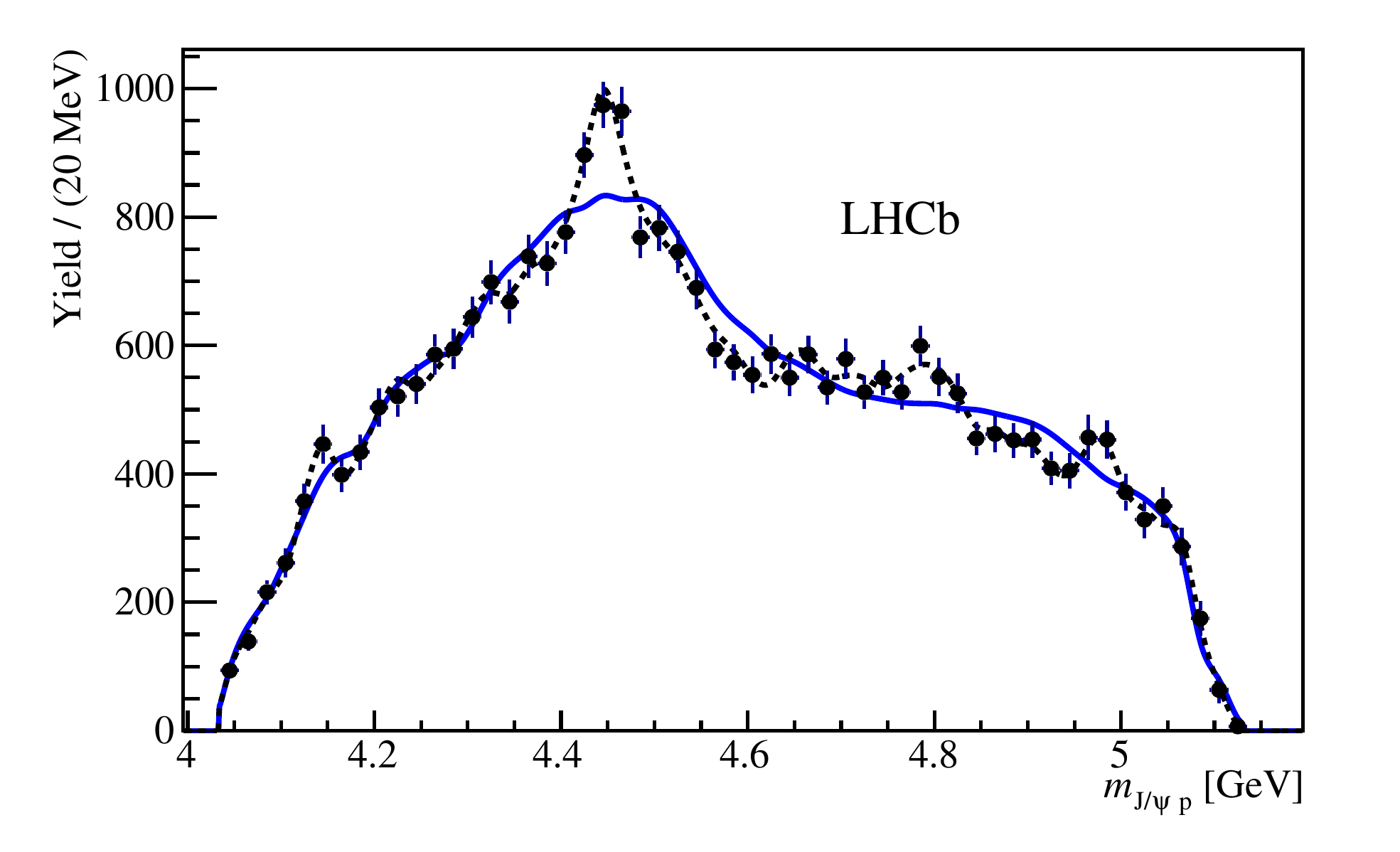}
  \end{center}
  \vskip-0.5cm\caption{\small 
    Efficiency-corrected and background-subtracted $m_{\jpsi p}$
    distribution 
    of the data (black points with error bars),  
    with $\PDF(m_{\jpsi p}|H_0)$ (solid blue line) and 
    $\PDF(m_{\jpsi p}|H_1)$ (dashed black line)
    superimposed.
   \label{fig:mjpsip}
  }
   \vskip-0.2cm
\end{figure}

Instead of using the two-dimensional (2D) distribution of $(m_{Kp},\cos\theta_{\LambdaStar})$ 
to evaluate the consistency of the data with the $H_0$ hypothesis, 
now expressed by the $l\leq\lmax(m_{Kp})$ requirement, 
it is more convenient to use the $m_{\jpsi p}$ ($m_{\jpsi K}$) distribution, 
as any deviations from $H_0$ should appear in the mass region of 
potential pentaquark (tetraquark) resonances.
The projection of $\PDF(m_{Kp},\cos\theta_{\LambdaStar}|H_0)$ onto $m_{\jpsi p}$ involves 
replacing $\cos\theta_{\LambdaStar}$ with $m_{\jpsi p}$ and integrating over $m_{Kp}$. 
This integration is carried out numerically, by generating 
large numbers of simulated events uniformly distributed in $m_{Kp}$ and $\cos\theta_{\LambdaStar}$, 
calculating the corresponding value of $m_{\jpsi p}$, 
and then filling a histogram with $\PDF(m_{Kp},\cos\theta_{\LambdaStar}|H_0)$ as a weight. 
In Fig.~\ref{fig:mjpsip}, $\PDF({m_{\jpsi p}}|H_0)$ 
is compared to the directly obtained efficiency-corrected 
and background-subtracted $m_{\jpsi p}$ distribution in the data.

To probe the compatibility of $\PDF(m_{\jpsi p}|H_0)$ with the data,
a sensitive test can be constructed by making a specific alternative hypothesis
($H_1$).
Following the method discussed in Ref.~\cite{LHCb-PAPER-2015-038} 
$H_1$ is defined as $l\leq\llarge$, where $\llarge$ is 
not dependent on $m_{Kp}$ and large enough to
reproduce structures induced by $\jpsi p$ or $\jpsi K$ contributions.
The significance of the $\lmax(m_{Kp})\leq l\leq\llarge$ Legendre moments
is probed using the likelihood ratio test:
\begin{equation}
\dll=\sum_{i=1}^{n_{\rm cand}^{\rm sig}+n_{\rm cand}^{\rm side}} 
w_i \ln \frac{ \PDF({m_{\jpsi p}}^i|H_0)/I_{H_{0}} }{ \PDF({m_{\jpsi p}}^i|H_1)/I_{H_{1}} } ,
\notag
\end{equation}
with normalizations $I_{H_{0,1}}$
determined via Monte Carlo integration. 
Note that the explicit event-by-event efficiency factor cancels in the likelihood ratio, 
but enters the likelihood normalizations.
In order for the test to have optimal sensitivity, the value $\llarge$
should be set such that the statistically significant features of
the data are properly described. Beyond that the power of the test
deteriorates. The limit $\llarge\to\infty$ would result in a perfect
description of the data, but a weak test since then the test statistic
would pick up the fluctuations in the data. For the same reason it is
also important to choose $\llarge$ independently of the actual data.
Here $\llarge=31$ is taken, one unit larger than the value used in
the model-independent analysis of $B^0\to\psi(2S)\pi^+K^-$ \cite{LHCb-PAPER-2015-038}, 
as baryons have half-integer spins.  
The result for $\PDF(m_{\jpsi p}|H_1)$ is shown in Fig.~\ref{fig:mjpsip}, where it is seen that $\llarge=31$ is sufficient. 
To make $\PDF(m_{\jpsi p}|H_{0,1})$ continuous, 
quadratic splines are used to interpolate between nearby $m_{\jpsi p}$ bins.

The numerical representations of $H_0$ and of $H_1$ contain 
a large number of parameters, 
requiring extensive statistical simulations to determine the distribution
of the test variable for the $H_0$ hypothesis: 
$\PDF_t(\dll|H_0)$.
A large number of pseudoexperiments are generated
with $n_{\rm cand}^{\rm sig}$ and $n_{\rm cand}^{\rm side}$ 
equal to those obtained in the data. 
The signal events, contributing a fraction $(1-\beta)$ to the signal region sample, 
are generated according to the $\PDF(m_{Kp},\cos\theta_{\LambdaStar}|H_0)$ function 
with parameters determined from the data. 
They are then shaped according to the $\epsilon(m_{Kp},\cos\theta_{\LambdaStar},\Omega_a)$ function, 
with the $\Omega_a$ angles generated uniformly in phase space.
The latter is an approximation, whose possible impact is discussed later.
Background events in sideband and signal regions are generated according to the 6D 
background parameterization previously developed in the amplitude analysis 
of the same data (Ref.~\cite{LHCb-PAPER-2015-029} supplement). 
The pseudoexperiments are subject to the same analysis procedure as the data. 
The distribution of values of $\dll$ over more than 10\,000 pseudoexperiments determines 
the form of $\PDF_t(\dll|H_0)$, 
which can then be used to convert the $\dll$ value obtained from data into a corresponding $p$-value. 
A small $p$-value indicates non-$\LambdaStar$ contributions in the data.
A large $p$-value means that the data are consistent with the $\LambdaStar$-only hypothesis, 
but does not rule out other contributions. 

\begin{figure}[hbtp]
   \vskip-0.3cm
  \begin{center}
  \includegraphics*[width=1.1\figsize]{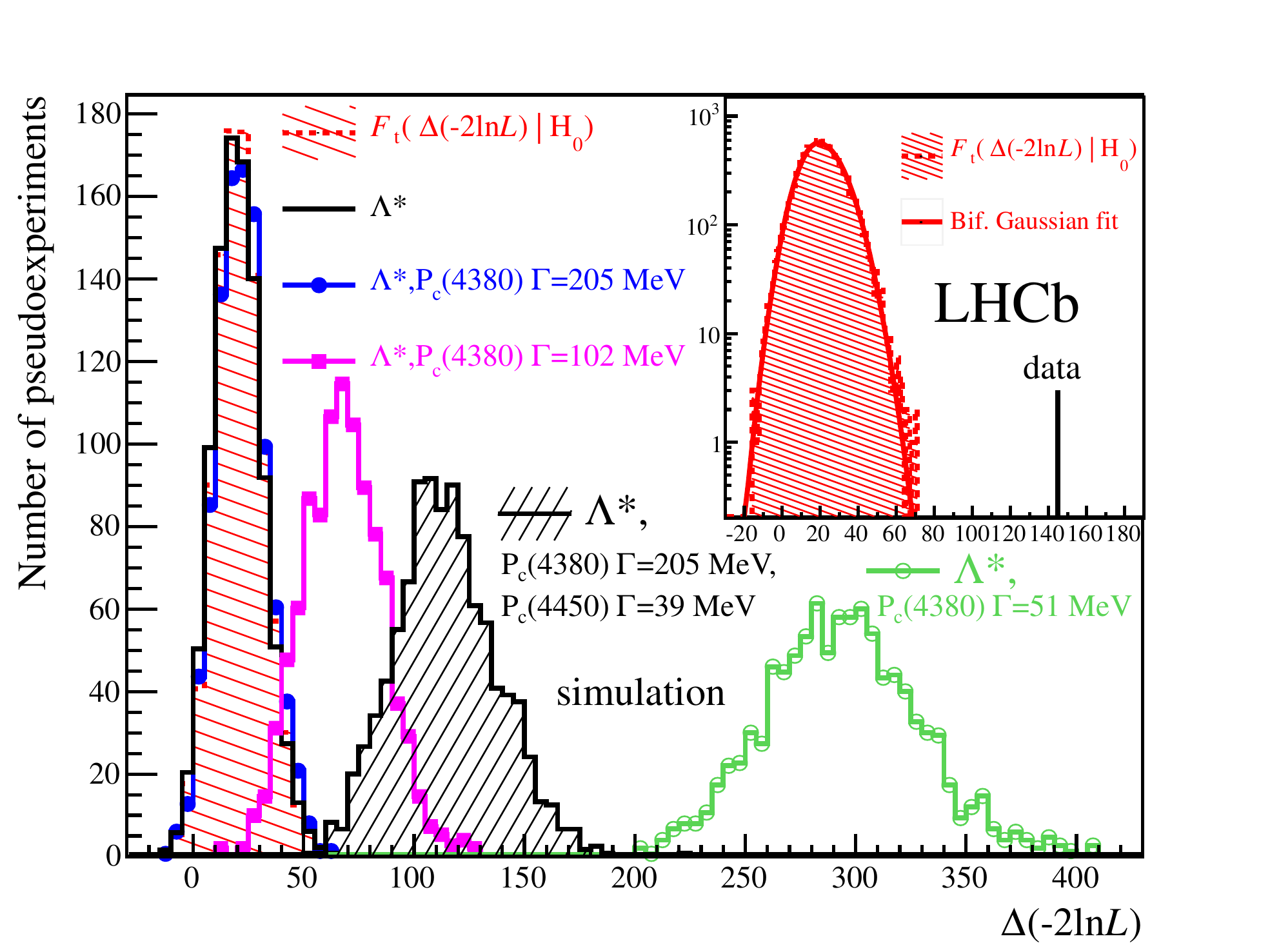} 
  \end{center}
  \vskip-0.3cm\caption{\small 
   Distributions of $\dll$ 
   in the model-independent pseudoexperiments 
   corresponding to $H_0$ (red falling hatched)  
   compared to the distributions for pseudoexperiments 
   generated from various amplitude models and, in the inset,
   to the bifurcated Gaussian fit function (solid line) and
   the value obtained for the data (vertical bar).
   \label{fig:dll}
  }
   \vskip-0.2cm
\end{figure}

Before applying this method to the data, 
it is useful to study its sensitivity with the help of amplitude models. 
Pseudoexperiments are generated according to 
the 6D amplitude model containing only $\LambdaStar$ resonances 
(the reduced model in Table~1 of Ref.~\cite{LHCb-PAPER-2015-029}), along with efficiency effects. 
The distribution of $\dll$ values 
is close to that expected from $\PDF_t(\dll|H_0)$ (black open and red falling hatched histograms in Fig.~\ref{fig:dll}),
thus verifying the 2D model-independent procedure on one example of the $\LambdaStar$ model. 
They also indicate that the non-uniformities in $\epsilon(\Omega_a)$ 
are small enough not to significantly bias the $\PDF_t(\dll|H_0)$ distribution 
when approximating the $\Omega_a$ probability density via a uniform distribution. 
To test the sensitivity of the method to an exotic $P_c^+\to\jpsi p$ resonance, 
the amplitude model described in Ref.~\cite{LHCb-PAPER-2015-029} is used, but
with the $P_c(4450)^+$ contribution removed.
Generating many pseudoexperiments from this amplitude model produces a distribution of $\dll$, 
which is almost indistinguishable from the $\PDF_t(\dll|H_0)$ distribution 
(blue dotted and red falling hatched histograms in Fig.~\ref{fig:dll}), 
thus predicting that for such a broad $P_c(4380)^+$ resonance ($\Gamma_0=205$ \mev) 
the false $H_0$ hypothesis is expected to be accepted (type II error),
because the $P_c(4380)^+$ contribution inevitably feeds into the numerical representation of $H_0$. 
Simulations are then repeated while reducing the $P_c(4380)^+$ width by subsequent factors of two,
showing a dramatic increase in the power of the test (histograms peaking at 60 and 300). 
Figure~\ref{fig:dll} also shows
the $\dll$ distribution obtained with the narrow $P_c(4450)^+$ 
state restored in the amplitude model and $P_c(4380)^+$ at its nominal $205$ \mev width
(black rising hatched histogram). 
The separation from $\PDF_t(\dll|H_0)$ is smaller than that of the simulation with
a $P_c(4380)^+$ 
of comparable width ($51\mev$)
due to 
the smaller $P_c(4450)^+$ fit fraction.
Nevertheless, the separation from $\PDF_t(\dll|H_0)$ is clear; thus,
if this amplitude model is a good representation of the data, 
the $H_0$ hypothesis is expected to essentially always be rejected. 

The value of the $\dll$ test variable obtained from the data 
is significantly above 
the $\PDF_t(\dll|H_0)$ distribution
(see the inset of Fig.~\ref{fig:dll}). 
To estimate a $p$-value  
the simulated $\PDF_t(\dll|H_0)$ 
distribution is fitted with a bifurcated Gaussian function (asymmetric widths);     
the significance of the $H_0$ rejection is $10.1\,\sigma$ standard deviations.

To test the sensitivity of the result to possible biases from the background 
subtraction, either the left 
or the right 
sideband is exclusively used,
and the weakest obtained rejection of $H_0$ is $9.8\,\sigma$. 
As a further check, the sideband subtraction is performed with
the \sPlot technique~\cite{2005NIMPA.555..356P},
in which the $w_i$ weights are obtained from the fit to the $m_{\jpsi p K}$ distribution
for candidates in the entire fit range. 
This increases the significance of the $H_0$ rejection to $10.4\,\sigma$.
Loosening the cut on the boosted decision tree variable discussed in Ref.~\cite{LHCb-PAPER-2015-029}
increases the signal efficiency by 14\%, while doubling the background fraction $\beta$,
and causes the significance of the $H_0$ rejection to increase to $11.1\,\sigma$. 
Replacing the uniform generation of the $\Omega_a$ angles in the $H_0$ pseudoexperiments with that of
the amplitude model without the $P_c(4380)^+$ and $P_c(4450)^+$ states, but generating $(m_{Kp},\cos\theta_{\LambdaStar})$ 
in the model-independent way, results in a $9.9\,\sigma$ $H_0$ rejection. 

\begin{figure}[tbhp]
   \vskip-0.2cm
  \begin{center}
  \includegraphics*[width=1.05\figsize]{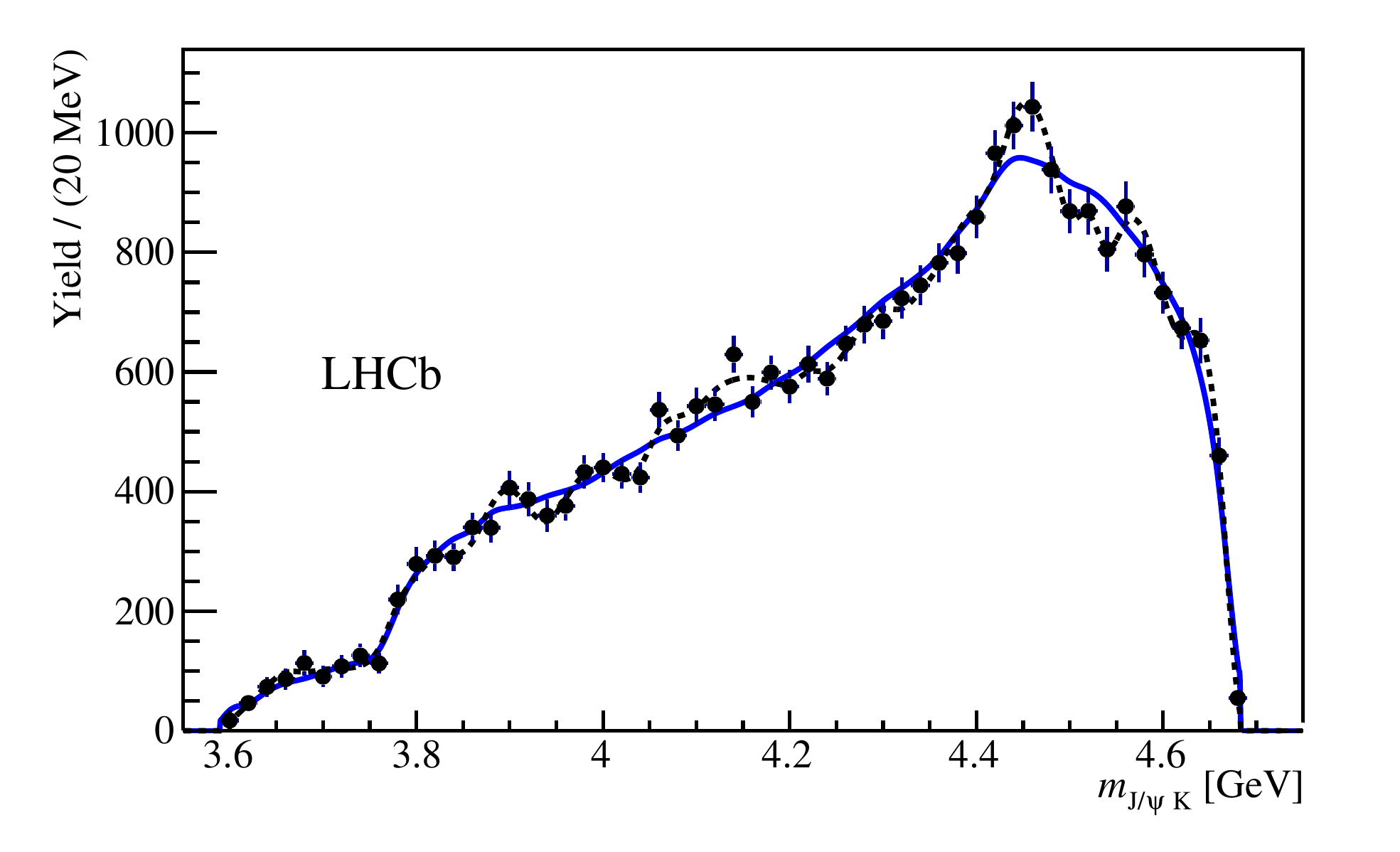}
  \end{center}
  \vskip-0.5cm\caption{\small 
    Efficiency-corrected and background-subtracted $m_{\jpsi K}$
    distribution 
    of the data (black points with error bars),  
    with $\PDF(m_{\jpsi K}|H_0)$ (solid blue line) and 
    $\PDF(m_{\jpsi K}|H_1)$ (dashed black line)
    superimposed.  
   \label{fig:mjpsik}
  }
   \vskip-0.1cm
\end{figure}
Figure~\ref{fig:mjpsip} indicates that the rejection of the $H_0$ hypothesis has to do with a narrow peak in the data near $4450$ \mev.
Determination of any $P_c^+$ parameters is not possible without a model-dependent analysis, 
because $P_c^+$ states feed into the numerical representation of $H_0$ in an intractable manner.

The $H_0$ testing is repeated using $m_{\jpsi K}$ instead of $m_{\jpsi p}$. 
The $m_{\jpsi K}$ distribution, with $\PDF(m_{\jpsi K}|H_0)$ and $\PDF(m_{\jpsi K}|H_1)$ superimposed, is shown in Fig.~\ref{fig:mjpsik}. 
The $\dll$ test gives a $5.3\,\sigma$ rejection of $H_0$,
which is lower than the rejection obtained using $m_{\jpsi p}$, thus providing 
model-independent evidence that non-$\LambdaStar$ contributions are more likely of the $P_c^+\to\jpsi p$ type. 
Further, in the model-dependent amplitude analysis \cite{LHCb-PAPER-2015-029}, it was seen that the $P_c$ states
reflected into the $m_{\jpsi K}$ distribution in the region in which $\PDF(m_{\jpsi K}|H_0)$ disagrees with the data.

In summary, it has been demonstrated at more than $9$ standard deviations that the $\Lb\to\jpsi p K^-$ 
decays cannot all be attributed to $K^-p$ resonant or nonresonant 
contributions. 
The analysis requires only minimal assumptions on the mass and spin of the $K^-p$ contributions; 
no assumptions on their number, their resonant or nonresonant nature, or their lineshapes have been made.
Non-$K^-p$ contributions, which must be present in the data, 
can be either of the exotic hadron type, or due to rescattering effects among ordinary hadrons.
This result supports the amplitude model-dependent observation 
of the $\jpsi p$ resonances presented previously \cite{LHCb-PAPER-2015-029}.

\quad\newline

\noindent We express our gratitude to our colleagues in the CERN
accelerator departments for the excellent performance of the LHC. We
thank the technical and administrative staff at the LHCb
institutes. We acknowledge support from CERN and from the national
agencies: CAPES, CNPq, FAPERJ and FINEP (Brazil); NSFC (China);
CNRS/IN2P3 (France); BMBF, DFG and MPG (Germany); INFN (Italy); 
FOM and NWO (The Netherlands); MNiSW and NCN (Poland); MEN/IFA (Romania); 
MinES and FANO (Russia); MinECo (Spain); SNSF and SER (Switzerland); 
NASU (Ukraine); STFC (United Kingdom); NSF (USA).
We acknowledge the computing resources that are provided by CERN, IN2P3 (France), KIT and DESY (Germany), INFN (Italy), SURF (The Netherlands), PIC (Spain), GridPP (United Kingdom), RRCKI and Yandex LLC (Russia), CSCS (Switzerland), IFIN-HH (Romania), CBPF (Brazil), PL-GRID (Poland) and OSC (USA). We are indebted to the communities behind the multiple open 
source software packages on which we depend.
Individual groups or members have received support from AvH Foundation (Germany),
EPLANET, Marie Sk\l{}odowska-Curie Actions and ERC (European Union), 
Conseil G\'{e}n\'{e}ral de Haute-Savoie, Labex ENIGMASS and OCEVU, 
R\'{e}gion Auvergne (France), RFBR and Yandex LLC (Russia), GVA, XuntaGal and GENCAT (Spain), Herchel Smith Fund, The Royal Society, Royal Commission for the Exhibition of 1851 and the Leverhulme Trust (United Kingdom).

\bibliographystyle{LHCb}
\bibliography{main,LHCb-PAPER,LHCb-CONF,LHCb-DP}

\newpage
\ifthenelse{\boolean{prl}}{\clearpage}{} 

\ifthenelse{\boolean{prl}}{}{ 
\section*{Appendix: Supplemental material}

\ifthenelse{\boolean{supp}}{\tableofcontents}{}

\section{Data sample}

The definition of the signal and sideband regions is illustrated in Fig.~\ref{fig:mjpsipk}. The background-subtracted and efficiency-corrected distribution of the data on the rectangular Dalitz plane $(m_{Kp},\cos\theta_{\LambdaStar})$ is shown in Fig.~\ref{fig:rectDalitz}.

\begin{figure}[tbhp]
\begin{center}
\includegraphics[width=\figsize]{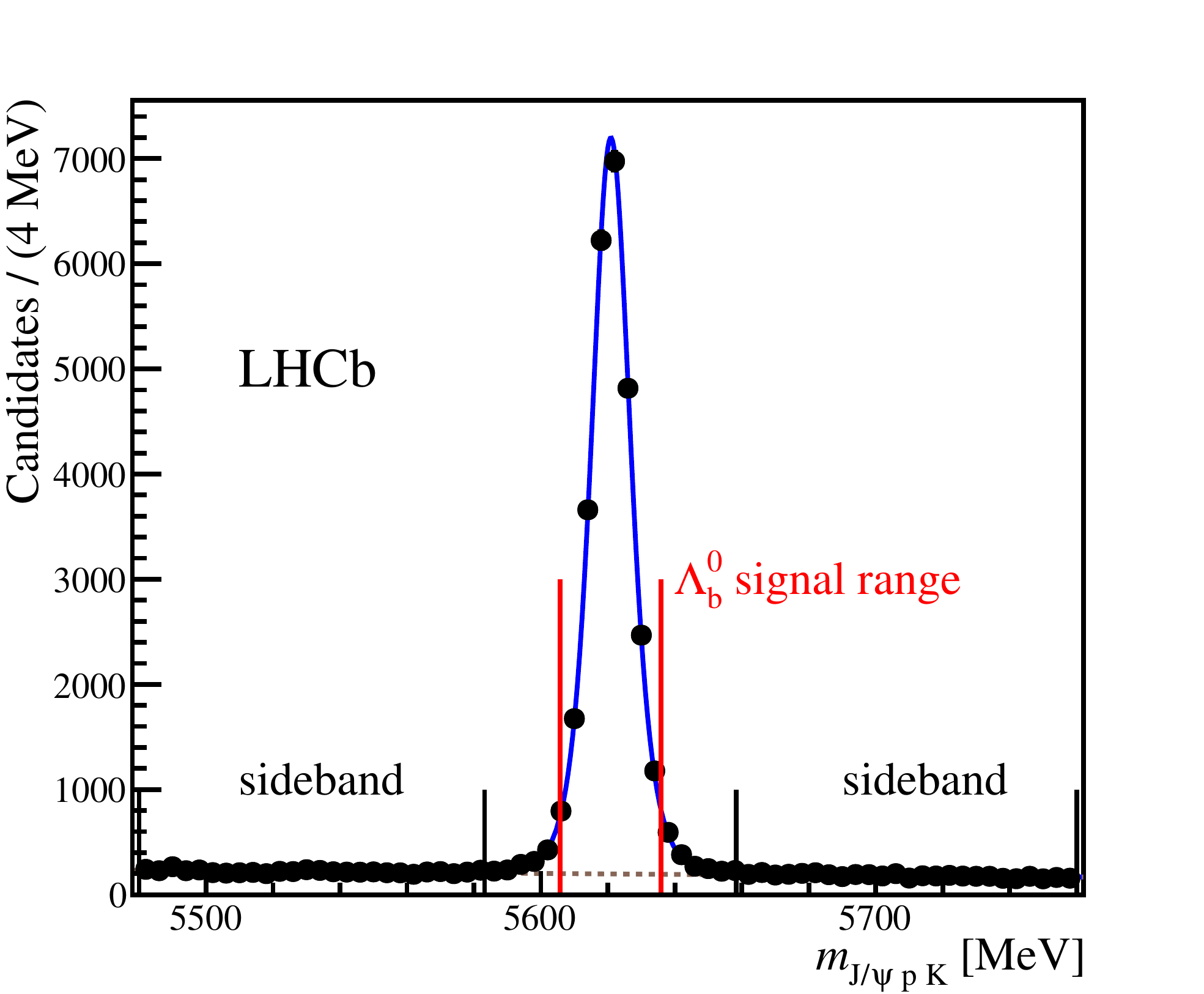}
\end{center}
\vskip -0.5cm
\caption{Distribution of $m_{\jpsi p K}$  in the data
with the fit of signal and background components superimposed 
\cite{LHCb-PAPER-2015-029}.
The fit is used to determine the background fraction $\beta$ 
in the $\pm2\sigma$ signal region
around the $\Lb$ peak (shown by the vertical red bars).
The sidebands used in the background subtraction are also shown. 
\label{fig:mjpsipk}
} 
\end{figure}

\begin{figure}[bthp]
\begin{center}
  \includegraphics[width=\figsize]{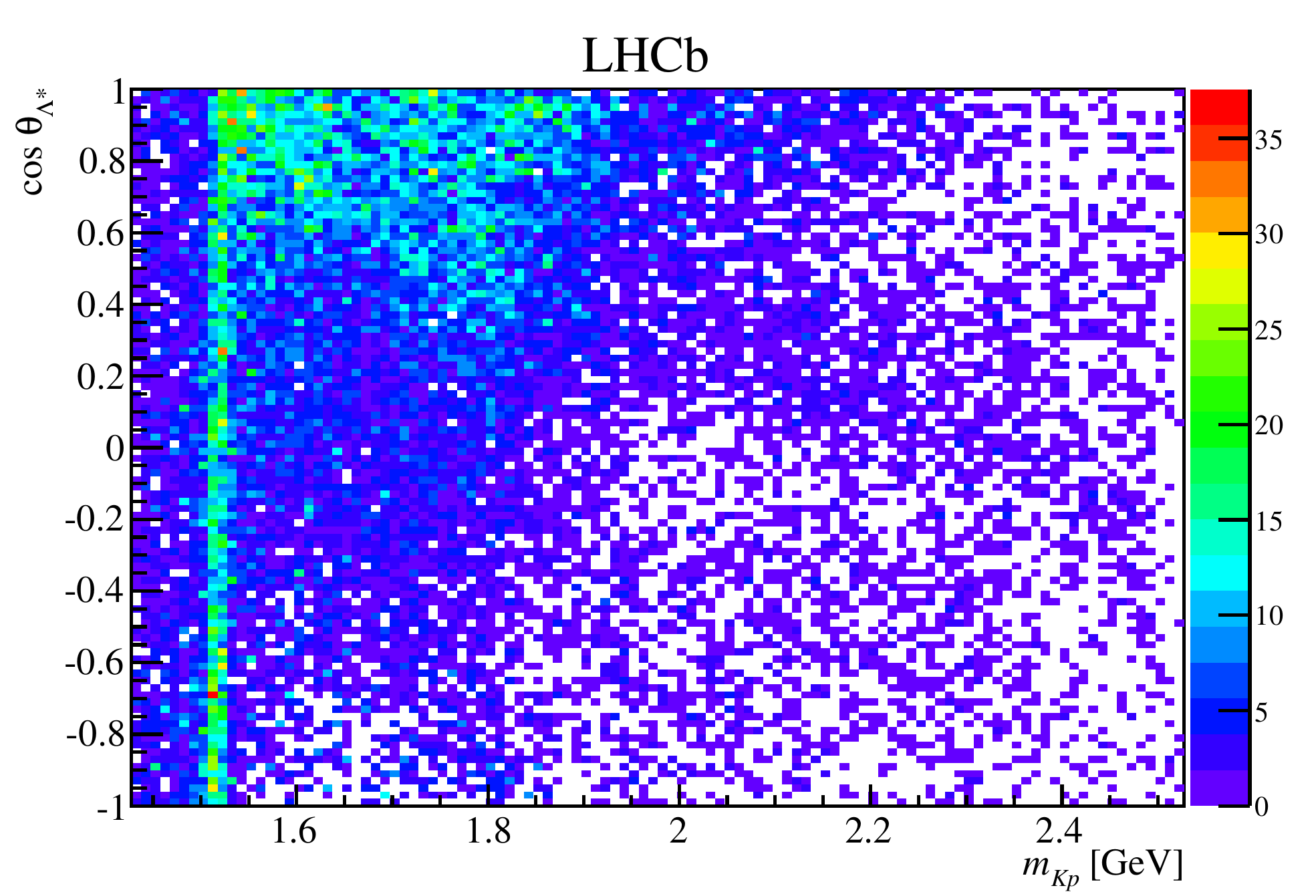}
\end{center}
\vskip -0.5cm
\caption{
Background-subtracted and efficiency-corrected distribution
of the cosine of the $\LambdaStar$ helicity angle versus $m_{Kp}$
for the data.
\label{fig:rectDalitz}
}
\end{figure}

\section{Simulations based on amplitude models}

The rectangular Dalitz plane $(m_{Kp},\cos\theta_{\LambdaStar})$ 
distributions 
for the large statistics pseudo-samples 
generated from the amplitude model with only the $\LambdaStar$ 
resonances 
and from the amplitude model with only the 
$P_c(4380)^+$ and $P_c(4450)^+$ resonances 
are shown in Figs.~\ref{fig:rectDalitzLz} 
and \ref{fig:rectDalitzTwoPc}, respectively. 
Parameters of the models, without and with the $P_c^+$ states,
were determined by fitting the amplitude models
to the data as described in Ref.~\cite{LHCb-PAPER-2015-029}.

The Legendre moments of $\cos\theta_{\LambdaStar}$ distributions
($\langle P_{l}^U \rangle^k$) 
in various bins of $m_{Kp}$ 
are compared between these two simulated pseudo-samples 
in Fig.~\ref{fig:momentsToys}. 
The $l\le\lmax(m_{Kp})$ filter, used in forming a numerical representation of
the hypothesis that only $K^-p$ contributions are present ($H_0$), 
is also illustrated in Fig.~\ref{fig:momentsToys}:
moments in the shaded regions ($l>\lmax(m_{Kp})$) are neglected. 
The pentaquark resonances can induce significant values of the moments
in these regions, as illustrated with the example amplitude model
containing only $P_c^+$ states. 
The $P_c^+$ states also contribute significantly 
to the unshaded $l\le\lmax(m_{Kp})$ regions,
thus feeding into the numerical representation of the $H_0$ hypothesis,
and decreasing the sensitivity of the model-independent approach to
exotic hadron contributions. 
This is especially true for wide resonances, which contribute
very little to high moments, as illustrated for the $P_c(4380)^+$ state 
in Fig.~\ref{fig:indpc}.
The example amplitude model with only $\LambdaStar$ resonances
contributes to the unshaded regions only, as expected.     

\begin{figure}[tbhp]
\quad\vskip-6.5cm
\begin{center}
  \includegraphics[width=\figsize]{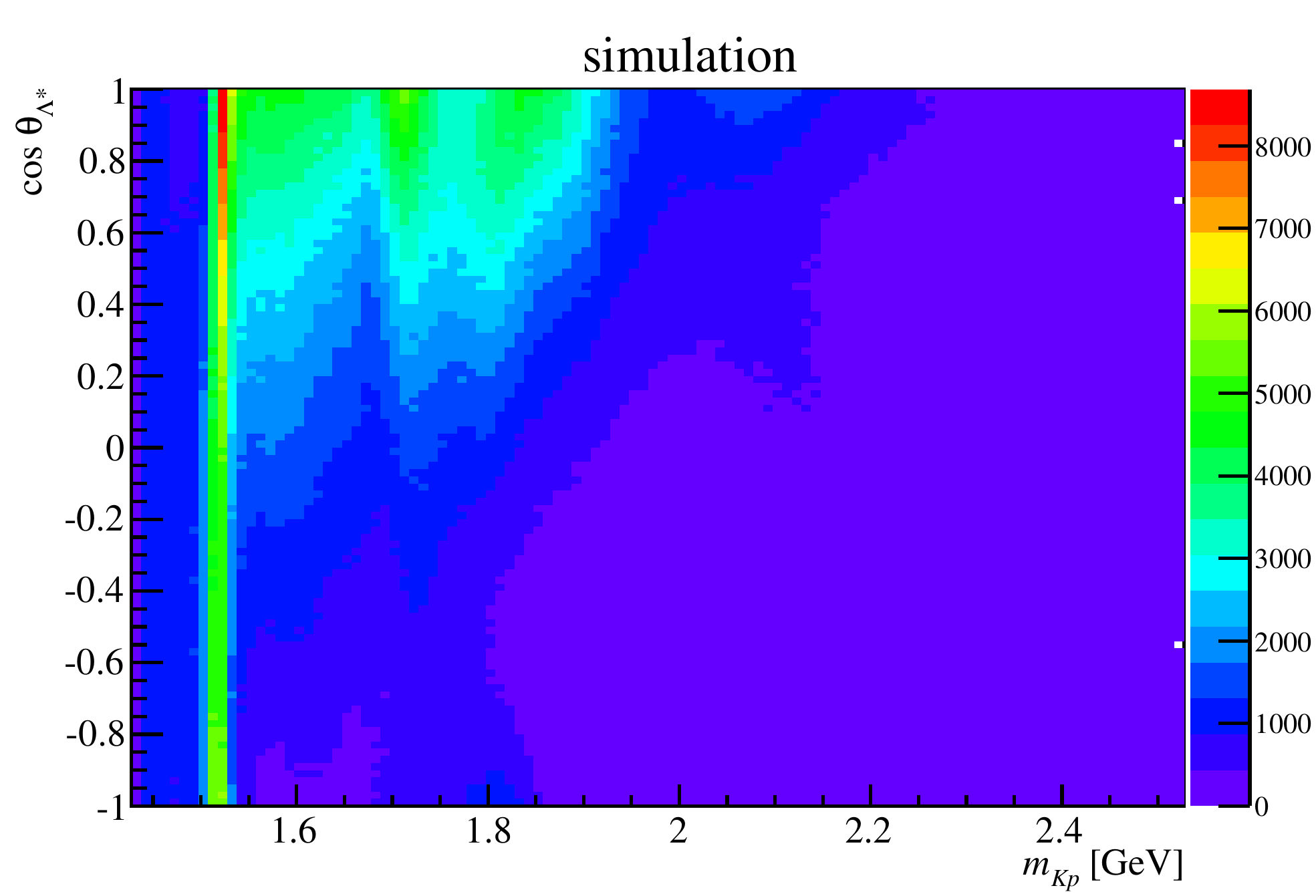}
\end{center}
\vskip -0.7cm
\caption{
Distribution in a pseudoexperiment
of the cosine of the $\LambdaStar$ helicity angle versus $m_{Kp}$
for the amplitude model with $\LambdaStar$ resonances only.
\label{fig:rectDalitzLz}
}
\vskip-1.0cm\quad
\end{figure}

\begin{figure}[bthp]
\begin{center}
  \includegraphics[width=\figsize]{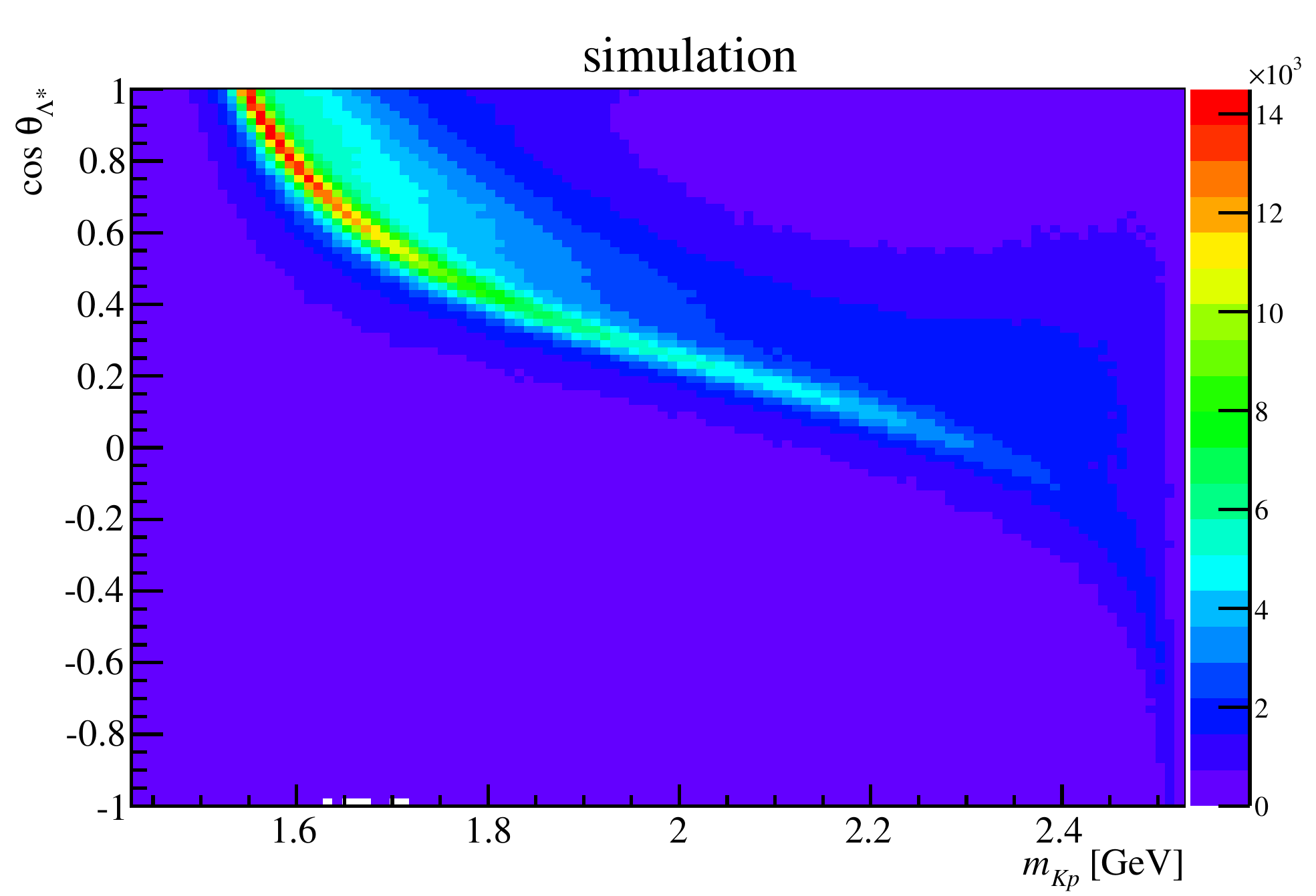}
\end{center}
\vskip -0.7cm
\caption{
Distribution in a pseudoexperiment
of the cosine of the $\LambdaStar$ helicity angle versus $m_{Kp}$
for the amplitude model with the $P_c(4380)^+$ and $P_c(4450)^+$ 
resonances only. 
\label{fig:rectDalitzTwoPc}
}
\vskip-1.0cm\quad
\end{figure}

\begin{figure}[tbhp]
\vbox{
\begin{center}
  \includegraphics[width=\figsize]{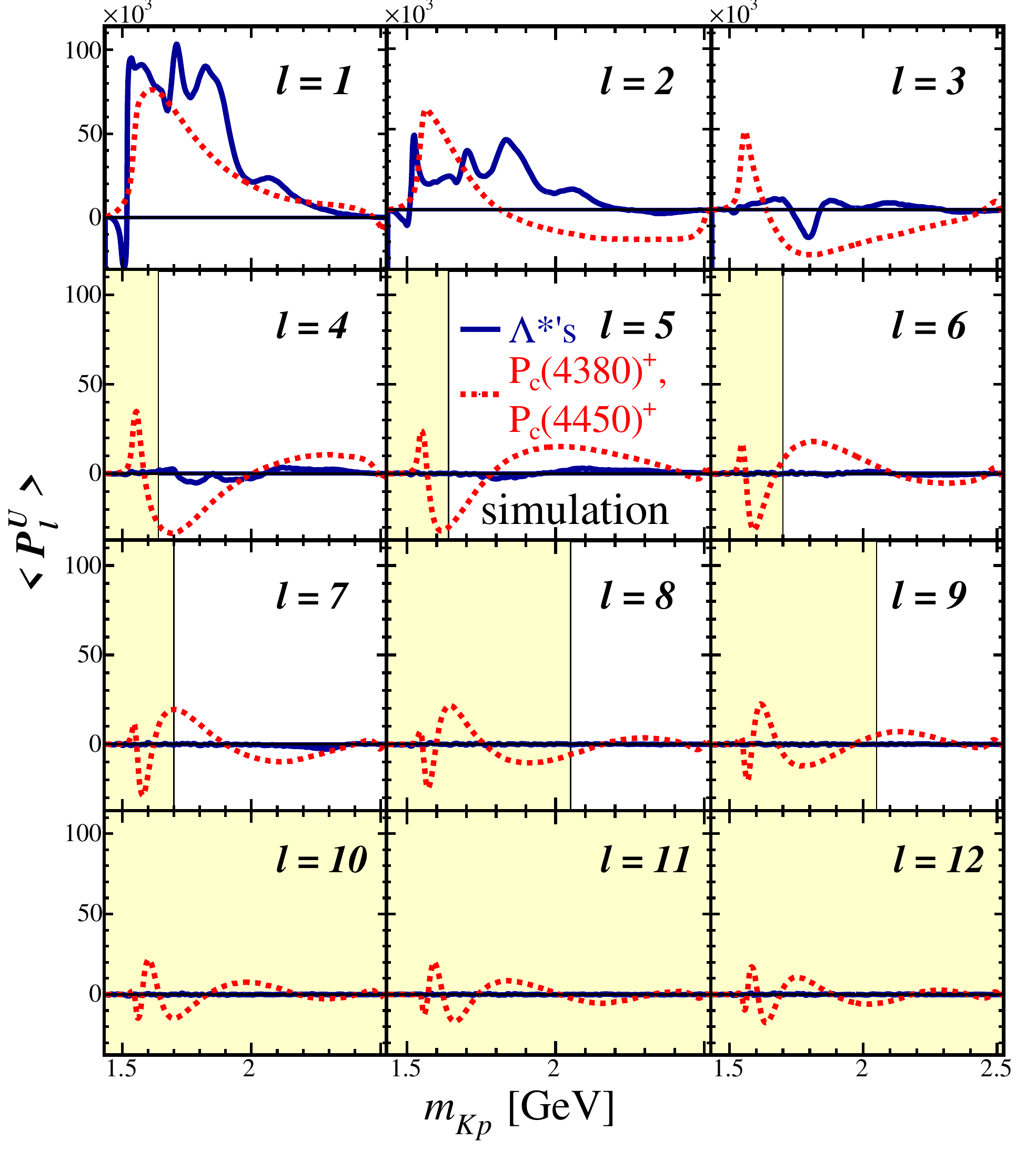}
\end{center}
\vskip-0.5cm\caption{\small 
           Legendre moments of $\cos\theta_{\LambdaStar}$ as a function of $m_{Kp}$ for 
           the simulated data from the amplitude models with only $\LambdaStar$ (solid blue lines)
           and with only $P_c(4380)^+$, $P_c(4450)^+$ contributions (dashed red lines), scaled by $0.5$.
           The regions excluded by the $l\le\lmax(m_{Kp})$ filter are shaded. 
\label{fig:momentsToys}
}
}
\end{figure}

\begin{figure}[tbhp]
\vbox{
\begin{center}
  \includegraphics[width=\figsize]{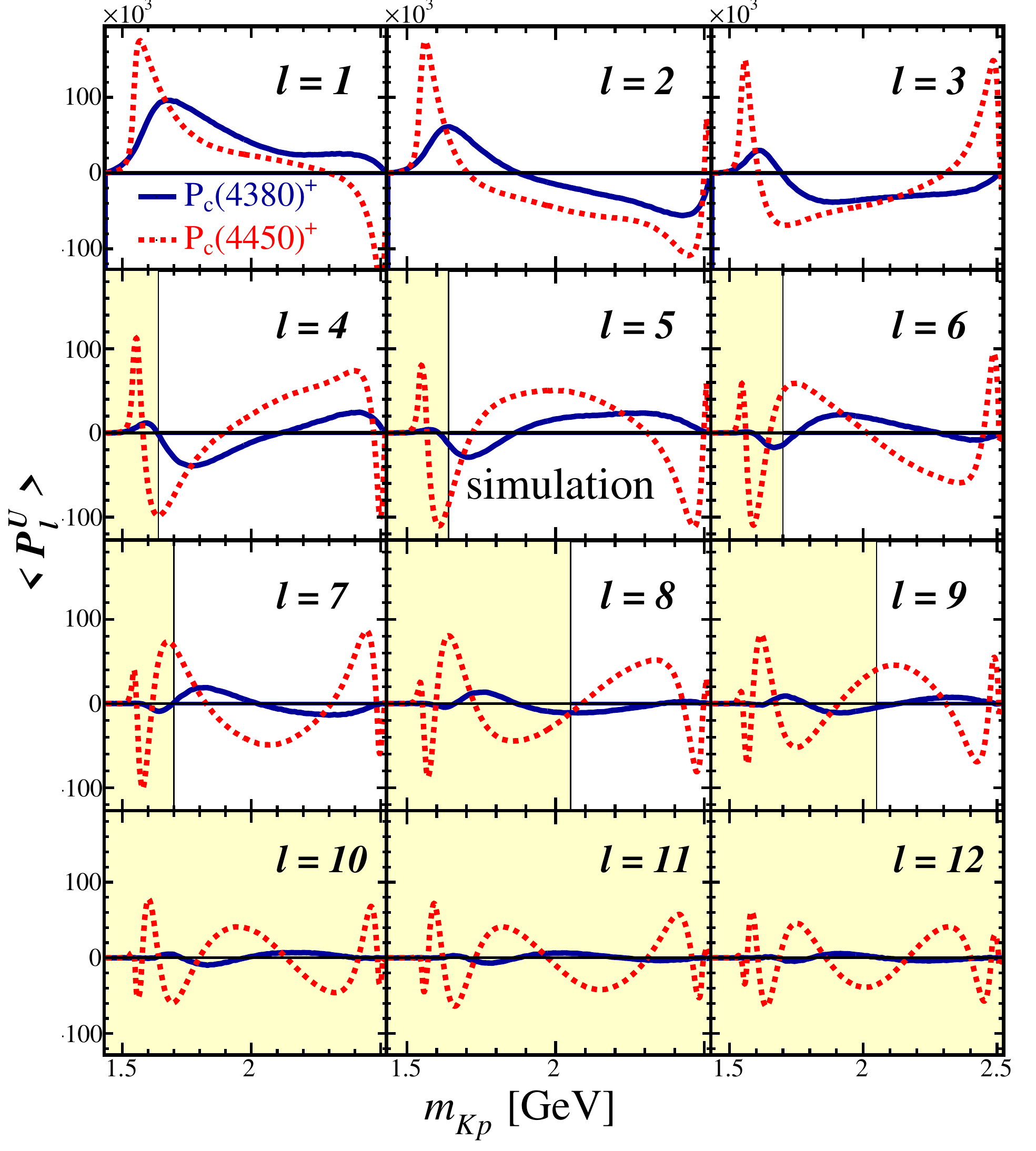}
\end{center}
\vskip-0.5cm\caption{\small 
           Legendre moments of $\cos\theta_{\LambdaStar}$ as a function of $m_{Kp}$ for 
           the simulated data from amplitude models with only $P_c(4380)^+$ (solid blue lines)
           and only $P_c(4450)^+$ contributions (dashed red line). 
\label{fig:indpc}
}
}
\end{figure}

\ifthenelse{\boolean{supp}}{
\bibliographystyle{LHCb}
\bibliography{main,LHCb-PAPER,LHCb-CONF,LHCb-DP}

\ifx\mcitethebibliography\mciteundefinedmacro
\PackageError{LHCb.bst}{mciteplus.sty has not been loaded}
{This bibstyle requires the use of the mciteplus package.}\fi
\providecommand{\href}[2]{#2}
\begin{mcitethebibliography}{10}
\mciteSetBstSublistMode{n}
\mciteSetBstMaxWidthForm{subitem}{\alph{mcitesubitemcount})}
\mciteSetBstSublistLabelBeginEnd{\mcitemaxwidthsubitemform\space}
{\relax}{\relax}

\bibitem{GellMann:1964nj}
M.~Gell-Mann, \ifthenelse{\boolean{articletitles}}{\emph{{A schematic model of
  baryons and mesons}},
  }{}\href{http://dx.doi.org/10.1016/S0031-9163(64)92001-3}{Phys.\ Lett.\
  \textbf{8} (1964) 214}\relax
\mciteBstWouldAddEndPuncttrue
\mciteSetBstMidEndSepPunct{\mcitedefaultmidpunct}
{\mcitedefaultendpunct}{\mcitedefaultseppunct}\relax
\EndOfBibitem
\bibitem{Zweig:1964}
G.~Zweig, \ifthenelse{\boolean{articletitles}}{\emph{{An SU$_3$ model for
  strong interaction symmetry and its breaking}}, }{}
\newblock
  {\href{http://cds.cern.ch/record/352337/files/CERN-TH-401.pdf}{CERN-TH-401,
  1964}}\relax
\mciteBstWouldAddEndPuncttrue
\mciteSetBstMidEndSepPunct{\mcitedefaultmidpunct}
{\mcitedefaultendpunct}{\mcitedefaultseppunct}\relax
\EndOfBibitem
\bibitem{LHCb-PAPER-2015-029}
LHCb collaboration, R.~Aaij {\em et~al.},
  \ifthenelse{\boolean{articletitles}}{\emph{{Observation of $J/\psi p$
  resonances consistent with pentaquark states in $\Lambda_b^0\to J/\psi pK^-$
  decays}}, }{}\href{http://dx.doi.org/10.1103/PhysRevLett.115.072001}{Phys.\
  Rev.\ Lett.\  \textbf{115} (2015) 072001},
  \href{http://arxiv.org/abs/1507.03414}{{\tt arXiv:1507.03414}}\relax
\mciteBstWouldAddEndPuncttrue
\mciteSetBstMidEndSepPunct{\mcitedefaultmidpunct}
{\mcitedefaultendpunct}{\mcitedefaultseppunct}\relax
\EndOfBibitem
\bibitem{SPECRDMMS}
C.~Fernandez-Ramirez {\em et~al.},
  \ifthenelse{\boolean{articletitles}}{\emph{{Coupled-channel model for
  $\overline{K}N$ scattering in the resonant region}},
  }{}\href{http://arxiv.org/abs/1510.07065}{{\tt arXiv:1510.07065}}\relax
\mciteBstWouldAddEndPuncttrue
\mciteSetBstMidEndSepPunct{\mcitedefaultmidpunct}
{\mcitedefaultendpunct}{\mcitedefaultseppunct}\relax
\EndOfBibitem
\bibitem{SPECFG}
R.~N. Faustov and V.~O. Galkin, \ifthenelse{\boolean{articletitles}}{\emph{{
  Strange baryon spectroscopy in the relativistic quark model}},
  }{}\href{http://dx.doi.org/10.1103/PhysRevD.92.054005}{Phys.\ Rev.\
  \textbf{D92} (2015) 054005}, \href{http://arxiv.org/abs/1507.04530}{{\tt
  arXiv:1507.04530}}\relax
\mciteBstWouldAddEndPuncttrue
\mciteSetBstMidEndSepPunct{\mcitedefaultmidpunct}
{\mcitedefaultendpunct}{\mcitedefaultseppunct}\relax
\EndOfBibitem
\bibitem{SPECCI}
S.~Capstick and N.~Isgur, \ifthenelse{\boolean{articletitles}}{\emph{{Baryons
  in a relativized quark model with chromodynamics}},
  }{}\href{http://dx.doi.org/10.1103/PhysRevD.34.2809}{Phys.\ Rev.\
  \textbf{D34} (1986) 2809}\relax
\mciteBstWouldAddEndPuncttrue
\mciteSetBstMidEndSepPunct{\mcitedefaultmidpunct}
{\mcitedefaultendpunct}{\mcitedefaultseppunct}\relax
\EndOfBibitem
\bibitem{SPECLMP}
U.~Loring, B.~C. Metsch, and H.~R. Petry,
  \ifthenelse{\boolean{articletitles}}{\emph{{The light-baryon spectrum in a
  relativistic quark model with instanton-induced quark forces}},
  }{}\href{http://dx.doi.org/10.1007/s100500170105}{Eur.\ Phys.\ J.\
  \textbf{A10} (2001) 395}, \href{http://arxiv.org/abs/hep-ph/0103289}{{\tt
  arXiv:hep-ph/0103289}}\relax
\mciteBstWouldAddEndPuncttrue
\mciteSetBstMidEndSepPunct{\mcitedefaultmidpunct}
{\mcitedefaultendpunct}{\mcitedefaultseppunct}\relax
\EndOfBibitem
\bibitem{SPECMPS}
T.~Melde, W.~Plessas, and B.~Sengl,
  \ifthenelse{\boolean{articletitles}}{\emph{{Quark-model identification of
  baryon ground and resonant states}},
  }{}\href{http://dx.doi.org/10.1103/PhysRevD.77.114002}{Phys.\ Rev.\
  \textbf{D77} (2008) 114002}, \href{http://arxiv.org/abs/0806.1454}{{\tt
  arXiv:0806.1454}}\relax
\mciteBstWouldAddEndPuncttrue
\mciteSetBstMidEndSepPunct{\mcitedefaultmidpunct}
{\mcitedefaultendpunct}{\mcitedefaultseppunct}\relax
\EndOfBibitem
\bibitem{SPECSF}
E.~Santopinto and J.~Ferretti,
  \ifthenelse{\boolean{articletitles}}{\emph{{Strange and nonstrange baryon
  spectra in the relativistic interacting quark-diquark model with a G\"ursey
  and Radicati-inspired exchange interaction}},
  }{}\href{http://dx.doi.org/10.1103/PhysRevC.92.025202}{Phys.\ Rev.\
  \textbf{C92} (2015) 025202}, \href{http://arxiv.org/abs/1412.7571}{{\tt
  arXiv:1412.7571}}\relax
\mciteBstWouldAddEndPuncttrue
\mciteSetBstMidEndSepPunct{\mcitedefaultmidpunct}
{\mcitedefaultendpunct}{\mcitedefaultseppunct}\relax
\EndOfBibitem
\bibitem{SPECELMS}
G.~P. Engel, C.~B. Lang, D.~Mohler, and A.~Schaefer,
  \ifthenelse{\boolean{articletitles}}{\emph{{ QCD with two light dynamical
  chirally improved quarks: baryons}},
  }{}\href{http://dx.doi.org/10.1103/PhysRevD.87.074504}{Phys.\ Rev.\
  \textbf{D87} (2013) 074504}, \href{http://arxiv.org/abs/1301.4318}{{\tt
  arXiv:1301.4318}}\relax
\mciteBstWouldAddEndPuncttrue
\mciteSetBstMidEndSepPunct{\mcitedefaultmidpunct}
{\mcitedefaultendpunct}{\mcitedefaultseppunct}\relax
\EndOfBibitem
\bibitem{PDG2014}
Particle Data Group, K.~A. Olive {\em et~al.},
  \ifthenelse{\boolean{articletitles}}{\emph{{\href{http://pdg.lbl.gov/}{Review
  of particle physics}}},
  }{}\href{http://dx.doi.org/10.1088/1674-1137/38/9/090001}{Chin.\ Phys.\
  \textbf{C38} (2014) 090001}, {and 2015 update}\relax
\mciteBstWouldAddEndPuncttrue
\mciteSetBstMidEndSepPunct{\mcitedefaultmidpunct}
{\mcitedefaultendpunct}{\mcitedefaultseppunct}\relax
\EndOfBibitem
\bibitem{Aubert:2008aa}
BaBar collaboration, B.~Aubert {\em et~al.},
  \ifthenelse{\boolean{articletitles}}{\emph{{Search for the $Z(4430)^-$ at
  BaBar}}, }{}\href{http://dx.doi.org/10.1103/PhysRevD.79.112001}{Phys.\ Rev.\
  \textbf{D79} (2009) 112001}, \href{http://arxiv.org/abs/0811.0564}{{\tt
  arXiv:0811.0564}}\relax
\mciteBstWouldAddEndPuncttrue
\mciteSetBstMidEndSepPunct{\mcitedefaultmidpunct}
{\mcitedefaultendpunct}{\mcitedefaultseppunct}\relax
\EndOfBibitem
\bibitem{LHCb-PAPER-2015-038}
LHCb collaboration, R.~Aaij {\em et~al.},
  \ifthenelse{\boolean{articletitles}}{\emph{{A model-independent confirmation
  of the $Z(4430)^-$ state}},
  }{}\href{http://dx.doi.org/10.1103/PhysRevD.92.112009}{Phys.\ Rev.\
  \textbf{D92} (2015) 112009}, \href{http://arxiv.org/abs/1510.01951}{{\tt
  arXiv:1510.01951}}\relax
\mciteBstWouldAddEndPuncttrue
\mciteSetBstMidEndSepPunct{\mcitedefaultmidpunct}
{\mcitedefaultendpunct}{\mcitedefaultseppunct}\relax
\EndOfBibitem
\bibitem{Alves:2008zz}
LHCb collaboration, A.~A. Alves~Jr.\ {\em et~al.},
  \ifthenelse{\boolean{articletitles}}{\emph{{The \lhcb detector at the LHC}},
  }{}\href{http://dx.doi.org/10.1088/1748-0221/3/08/S08005}{JINST \textbf{3}
  (2008) S08005}\relax
\mciteBstWouldAddEndPuncttrue
\mciteSetBstMidEndSepPunct{\mcitedefaultmidpunct}
{\mcitedefaultendpunct}{\mcitedefaultseppunct}\relax
\EndOfBibitem
\bibitem{Donoghue:1979mu}
J.~F. Donoghue, E.~Golowich, W.~A. Ponce, and B.~R. Holstein,
  \ifthenelse{\boolean{articletitles}}{\emph{{Analysis of $\Delta$S=1
  nonleptonic weak decays and the $\Delta$I=1/2 rule}},
  }{}\href{http://dx.doi.org/10.1103/PhysRevD.21.186}{Phys.\ Rev.\
  \textbf{D21} (1980) 186}\relax
\mciteBstWouldAddEndPuncttrue
\mciteSetBstMidEndSepPunct{\mcitedefaultmidpunct}
{\mcitedefaultendpunct}{\mcitedefaultseppunct}\relax
\EndOfBibitem
\bibitem{2005NIMPA.555..356P}
M.~Pivk and F.~R. Le~Diberder,
  \ifthenelse{\boolean{articletitles}}{\emph{{sPlot: A statistical tool to
  unfold data distributions}},
  }{}\href{http://dx.doi.org/10.1016/j.nima.2005.08.106}{{Nucl.\ Instrum.\
  Meth.\ } \textbf{A555} (2005) 356},
  \href{http://arxiv.org/abs/physics/0402083v3}{{\tt
  arXiv:physics/0402083v3}}\relax
\mciteBstWouldAddEndPuncttrue
\mciteSetBstMidEndSepPunct{\mcitedefaultmidpunct}
{\mcitedefaultendpunct}{\mcitedefaultseppunct}\relax
\EndOfBibitem
\end{mcitethebibliography}
}

\ifthenelse{\boolean{prl}}{}{\clearpage}

}

\clearpage

\centerline{\large\bf LHCb collaboration}
\begin{flushleft}
\small
R.~Aaij$^{39}$, 
C.~Abell\'{a}n~Beteta$^{41}$, 
B.~Adeva$^{38}$, 
M.~Adinolfi$^{47}$, 
Z.~Ajaltouni$^{5}$, 
S.~Akar$^{6}$, 
J.~Albrecht$^{10}$, 
F.~Alessio$^{39}$, 
M.~Alexander$^{52}$, 
S.~Ali$^{42}$
% ...

{\footnotesize \it
% ...
$ ^{5}$Clermont Universit\'{e}, Universit\'{e} Blaise Pascal, CNRS/IN2P3, LPC, Clermont-Ferrand, France\\
$ ^{6}$CPPM, Aix-Marseille Universit\'{e}, CNRS/IN2P3, Marseille, France\\
% ...
$ ^{10}$Fakult\"{a}t Physik, Technische Universit\"{a}t Dortmund, Dortmund, Germany\\
% ...
$ ^{38}$Universidad de Santiago de Compostela, Santiago de Compostela, Spain\\
$ ^{39}$European Organization for Nuclear Research (CERN), Geneva, Switzerland\\
% ...
$ ^{41}$Physik-Institut, Universit\"{a}t Z\"{u}rich, Z\"{u}rich, Switzerland\\
$ ^{42}$Nikhef National Institute for Subatomic Physics, Amsterdam, The Netherlands\\
% ...
$ ^{47}$H.H. Wills Physics Laboratory, University of Bristol, Bristol, United Kingdom\\
% ...
$ ^{52}$School of Physics and Astronomy, University of Glasgow, Glasgow, United Kingdom\\
% ...
}
\end{flushleft}


\end{document}